%% file: main.tex
\documentclass[sigconf]{acmart}

\def\BibTeX{{\rm B\kern-.05em{\sc i\kern-.025em b}\kern-.08emT\kern-.1667em\lower.7ex\hbox{E}\kern-.125emX}}
    
\usepackage{geometry}
\usepackage{tabularx}
\usepackage{booktabs}
\usepackage{caption}
\usepackage{natbib}
\usepackage{fancyhdr}
\newcolumntype{C}{>{\centering\arraybackslash}X}
\begin{document}
\fancyhead{}
\title[SAIN: Self-Attentive Integration Network for Recommendation]{SAIN: Self-Attentive Integration Network for Recommendation}

\author{Seoungjun Yun$^{\text{1}}$, Raehyun Kim$^{\text{1}}$, Miyoung Ko, Jaewoo Kang*}\thanks{$^1$ equal contribution, * corresponding author}
\affiliation{Department of Computer Science and Engineering, Korea University}
\email{{ysj5419, raehyun, gomi1503, kangj}@korea.ac.kr}

%
\renewcommand{\shortauthors}{Yun and Kim et al}

%
\begin{abstract}
With the growing importance of personalized recommendation, numerous recommendation models have been proposed recently. Among them, Matrix Factorization (MF) based models are the most widely used in the recommendation field due to their high performance. However, MF based models suffer from cold start problems where user-item interactions are sparse. To deal with this problem, content based recommendation models which use the auxiliary attributes of users and items have been proposed. Since these models use auxiliary attributes, they are effective in cold start settings. However, most of the proposed models are either unable to capture complex feature interactions or not properly designed to combine user-item feedback information with content information. 
In this paper, we propose Self-Attentive Integration Network (SAIN) which is a model that effectively combines user-item feedback information and auxiliary information for recommendation task. In SAIN, a self-attention mechanism is used in the feature-level interaction layer to effectively consider interactions between multiple features, while the information integration layer adaptively combines content and feedback information. The experimental results on two public datasets show that our model outperforms the state-of-the-art models by 2.13\%. 
\end{abstract}
%

\copyrightyear{2019} 
\acmYear{2019} 
\setcopyright{acmcopyright}
\acmConference[SIGIR '19]{Proceedings of the 42nd International ACM SIGIR Conference on Research and Development in Information Retrieval}{July 21--25, 2019}{Paris, France}
\acmBooktitle{Proceedings of the 42nd International ACM SIGIR Conference on Research and Development in Information Retrieval (SIGIR '19), July 21--25, 2019, Paris, France}
\acmPrice{15.00}
\acmDOI{10.1145/3331184.3331342}
\acmISBN{978-1-4503-6172-9/19/07}

\maketitle

\input{sections/Introduction.tex}
\input{sections/Methods.tex}
\input{sections/Experiments.tex}
\input{sections/Conclusion.tex}

\section*{Acknowledgement}
 This work was supported by the National Research Foundation of Korea (NRF-2017R1A2A1A17069645, NRF-2017M3C4A7065887)
\vspace{-1pt}
%
\bibliographystyle{ACM-Reference-Format}
\bibliography{reference}
\end{document}

%% file: sections/Introduction.tex
\section{Introduction}\label{sec:intro}
With the growing importance of recommender systems, numerous models for highly personalized recommendation have been proposed. Latent Factor Models (LFMs) such as Matrix Factorization (MF) \cite{koren2009matrix} and Factorization Machine (FM) \cite{rendle2010factorization} are widely used. Latent factor models can learn user and item representations in an end-to-end fashion without any explicit feature engineering. 

For example, MF utilizes only user-item interaction data, assuming that similar items would be located close to each other in a latent space during the learning process. Based on this assumption, MF models with deep neural network have been applied to capture more complex user-item relations in a non-linear way\cite{he2017neural}. Despite their simple structures, MF models have proven to be effective in various recommendation tasks.

\begin{figure}[h]
    \begin{center}
    \includegraphics[width=7.5cm,height=4cm]{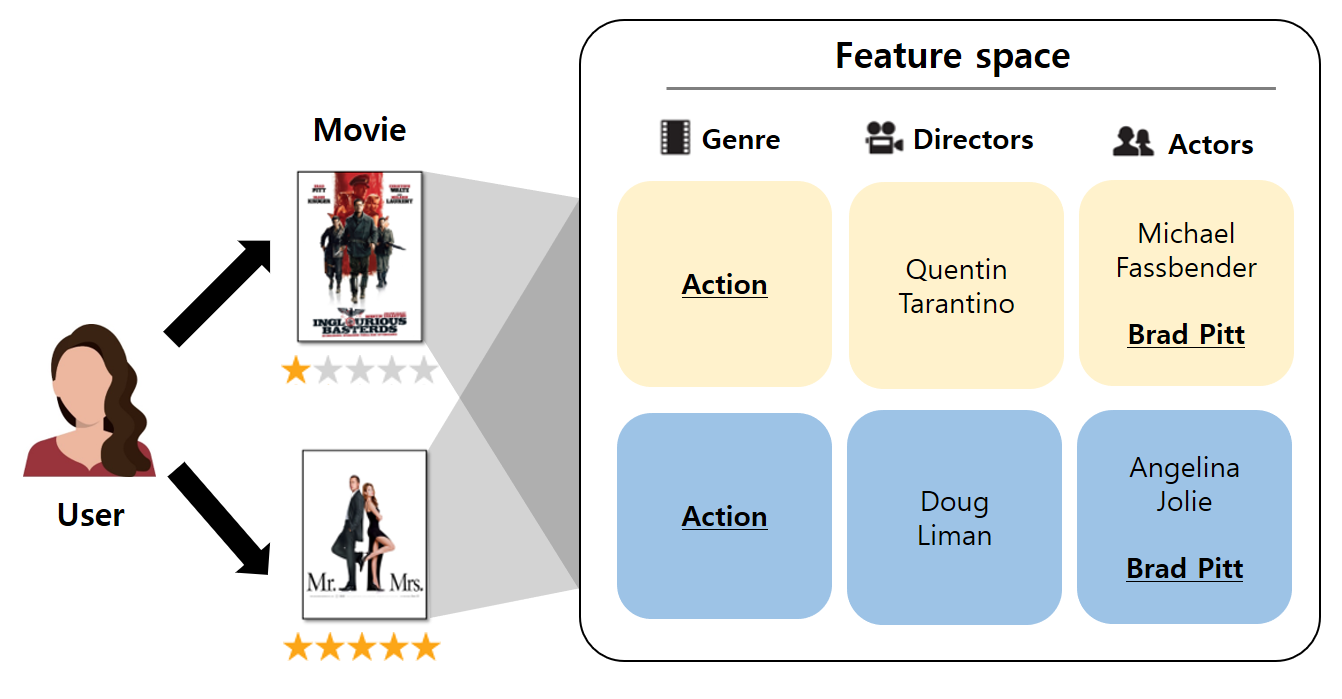}
    \caption{Two movies in the same genre but with different characteristics.}
    \label{fig:intro_ex}
    \end{center}
    \vspace{-4mm}
\end{figure}

However, since MF models utilize only user-item interaction data, they cannot effectively learn user and item representations when there is an insufficient amount of interaction data. If a user or an item is newly entered into the system or an item is not popular, there can be a lack of data. This problem is known as the cold-start problem. To overcome this limitation, models that incorporate auxiliary information have been proposed recently. For example, in \cite{chen2017attentive} and \cite{Chen2018heterogeneousattention}, multiple features are combined based on user preferences using attention mechanism. However, most of the proposed models are not designed to consider high order dependencies between features. The models assume that features are not affected by other features, which is not realistic. As illustrated in Figure \ref{fig:intro_ex}, the same user gives very different ratings to two movies in the same genre (Action) and with the same actor (Brad Pitt). The figure shows that the features of an item interact with each other to create a synergetic effect and determine the characteristics of the item. As determining which feature combinations are synergetic is very costly, we need a model that can automatically learn combinative feature representations. 

Factorization Machine (FM) models, which are another line of LFMs, are designed to consider feature interactions. FM models learn all features related to user and item entities as latent factors and predict ratings based on the second-order interactions between the features. FM models with deep neural networks are proposed to learn higher-order interactions between features and capture non-linear relationships between features \cite{he2017nfm,lian2018xdeepfm}. Although FM models have proven to be effective on sparse data, they are not properly designed to combine 
user-item feedback information with content information.
In \cite{liu2018context}, treating users, items and features as the same level entities may over-estimate the influence 
of content information, thus weakening the impacts of user-item feedback information. Side information might become a noise for a user with relatively large interactions. In that case, the model should give more weight to user-item feedback information.

To overcome the limitations of MF and FM, we propose Self-Attentive Integration Network (SAIN) in this paper. The contributions of this work can be summarized as follows. We designed a model that effectively combines user-item feedback information and content information to infer user preferences. To consider complex feature interactions, we adopted a Multi-head Self-attention network that can capture interactions between entities \cite{vaswani2017attention}. We conducted experiments on the public datasets MovieLens and Weibo-Douban. The experimental results show that our SAIN model outperforms the state-of-the-art MF and FM models in the rating prediction task. 

%% file: sections/Methods.tex
\section{Self-Attentive Integration Network}\label{sec:model_intro}
\begin{figure}[t]
    \begin{center}
    \includegraphics[width=8.2cm,height=5.3cm]{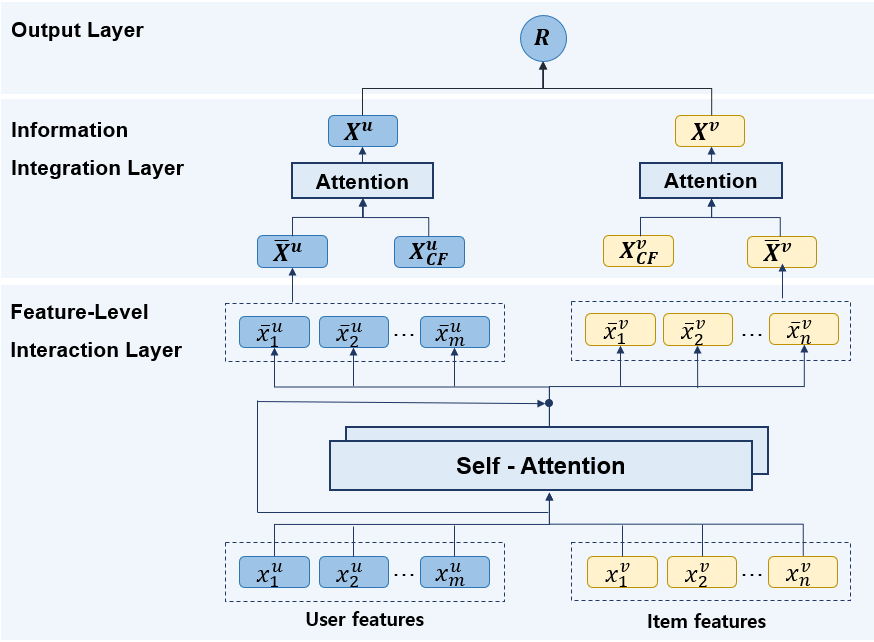}
    \caption{Self-Attentive Integration Network.}
    \label{fig:model structure}
    \end{center}
    \vspace{-2mm}
\end{figure}

In this section, we introduce our Self-Attentive Integration Network (SAIN) model which can be used for the rating prediction task. 
SAIN effectively combines user-item feedback information and content information to infer user preferences.
Figure \ref{fig:model structure} illustrates the overall architecture of SAIN.
The network is composed of the following three layers: 1) \textit{Feature-Level Interaction Layer} which generates user/item combinative feature representations while capturing high-order feature interactions, 2) \textit{Information Integration Layer} which combines user preference information from user-item feedback and content information from combinative feature representations, and 3) \textit{Output Layer} which predicts user ratings on items based on user/item representations.

\subsection{Feature-Level Interaction Layer}\label{subsec:feat_lev_layer}
The feature-level interaction layer learns how to effectively capture high-order interactions between multiple features to produce combinative feature representations. We adopted Multi-head Self-attention which is widely used for capturing interactions between entities. 
Let $u$ $\in$ $U$ and $v$ $\in$ $V$ denote a user and an item, respectively. Each user $u$ has a set of content features $\boldsymbol{x^u} = \{x_{1}^{u},x_{2}^{u},...,x_{m}^{u}\}$, where $m$ is the number of user content features. Similarly, each item $v$ has a set of content features $\boldsymbol{x^v} = \{x_{1}^{v},x_{2}^{v},...,x_{n}^{v}\}$, where $n$ is the number of item features. Each element of a content feature set represents a $d$-dimensional latent vector. To consider interactions between items and those between user features, we combine the set of user features and the set of item features and use the combined set $\boldsymbol{x} = \{\boldsymbol{x^u}, \boldsymbol{x^v}\}$ as input for our SAIN model.  

All latent vectors of each feature are converted to $d'$-dimensional query, key, and value vectors for each attention head. We then measure the interaction between the $i^{th}$ and $j^{th}$ features using scaled dot-product attention under a specific attention head $h$ as follows:
\begin{equation}
    \alpha_{ij}^{(h)} = \frac{\exp e_{ij}}{\sum_{k=1}^{n+m} \exp e_{ik}}\, , 
      \quad \; e_{ij} = \frac{(x_i W_{Q}^{(h)})(x_j W_{K}^{(h)})}{\sqrt{d'}}\, ,
\end{equation}
where $W_{Q}^{(h)}, W_{K}^{(h)} \in R^{d' \times d}$.
We used only features with the $k$ highest attention scores to filter less informative interactions and avoid overfitting. Each feature vector is obtained by summing all the weighted feature vectors with the $k$ highest attention scores:
\begin{equation}
    \Tilde{x}_i^{(h)} = \sum_{j=1}^k \alpha_{ij}(x_j W_{V}^{(h)})\, .
\end{equation}
Different feature representations are obtained from each attention head $h$. By projecting original latent vectors to multiple different subspaces, each head can capture various aspects in different subspaces. Feature vectors from all heads are concatenated as follows:
\begin{equation}
    \Tilde{x}_i = Concat(\Tilde{x}_i^{(1)},\Tilde{x}_i^{(2)} ,\cdot \cdot \cdot ,\Tilde{x}_i^{(H)})\, ,
\end{equation}
where H denotes the number of attention heads.
Although interactions between features provide us important information, we should also preserve features' original characteristics. We add original latent vectors $x_i$ to new vectors after multi-head attention $\Tilde{x}_i$ by residual connections as follows:
\begin{equation}
    \bar{x}_i = \max(\Tilde{x}_i + x_i, 0)\, ,
\end{equation}
where max($\cdot$,0) denotes the Rectifier (ReLU) activation function.
Given new representations of features $\bar{x}$, we map user feature representations $\bar{X}^u$ to latent vectors of user features, and we map item feature representations $\bar{X}^v$ to latent vectors of item features as follows:
\begin{equation}
    \bar{X}^u = f_{user}(Concat(\bar{x}_1^u,\bar{x}_2^u,\cdot \cdot \cdot, \bar{x}_m^u))\, ,
\end{equation}
\begin{equation}
    \bar{X}^v= f_{item}(Concat(\bar{x}_{1}^v,\bar{x}_{2}^v,\cdot \cdot \cdot, \bar{x}_{n}^v))\, ,
\end{equation}
where $f_{user}$, $f_{item}$ denotes a fully connected layer. 

We use the dot product to model the interactions between features as in the latent factor models. The scores with feature representations of users and items can be considered as the rating predictions based on the contents. The content score is calculated as follows: $Score_{contents} = \bar{X}^u \cdot \bar{X}^v $.

\subsection{Information Integration Layer}\label{subsec:info_integ_layer}
The information integration layer adaptively combines the feedback and content information of users and items using attention mechanism which has proven to be effective \cite{Chen2018heterogeneousattention,Shi2018attentioncold}. The importance of each type of information varies according to the circumstances. If a system has a considerable amount of user interaction data, we should give more weight to feedback information. On the other hand, if a user or item is relatively new in the system, we should give more weight to feature information.

 We use the outputs of the feature-level interaction layer as content representations. To obtain information from user-item feedback, we train an independent feedback latent vector. The preference score is calculated as $Score_{pr} = X^u_{CF} \cdot X^v_{CF}\, ,$ where $X^u_{CF}$ is a user feedback latent vector and $X^v_{CF}$ is an item feedback latent vector. We used attention mechanism to combine the two latent vectors as follows:

\begin{equation}
    \alpha_{CF} = \frac{exp(g(X_{CF}))}{exp(g(X_{CF})+exp(g(\bar {X}))}\, ,
\end{equation}
where $g$ denotes a fully connected layer.
By combining the vectors using Eq. (\ref{eq:combination}), we obtain the final user and item representations. 
\begin{equation}
    X = \alpha_{CF} * X_{CF} + (1-\alpha_{CF})*\bar {X}
    \label{eq:combination}
\end{equation}

\subsection{Output Layer}
Our final prediction is based on the user and item representations from the information integration layer. The combined score is calculated as $Score_{combined} = X^u \cdot X^v$ where $X^u$ and $X^v$ are user and item combined latent vectors (feature representation vector + feedback latent vector), respectively. We used Root Mean Squared Error (RMSE) as our metric and reported the Mean Absolute Error (MAE) scores in the Experiments section \ref{subsec:perf_comp}. 
\begin{equation}
    RMSE = \sqrt{\frac{1}{N}\sum_i^N{(Rating_i - Score_i)^2}}
    \label{eq:RMSE}
\end{equation}
We trained our parameters to minimize the RMSE for all three scores (Content, Preference, Combined). By jointly minimizing all three losses, feedback and feature latent vectors are trained independently so they can be used to obtain content and feedback information. Our model finds ways to effectively integrate the two types of information. In the evaluation stage, we used only the combined score to predict ratings. 

%% file: sections/Experiments.tex
\section{Experiments}
In this section, we compare our proposed SAIN model with various baseline models to show the effectiveness of our model.
\subsection{Datasets}
We conducted experiments on two public datasets. As our work focuses on the importance of interactions between multiple features, we picked datasets with user and item features. The statistics of the datasets are summarized in Table 1. The number of ratings ranges from 1 to 5 in both datasets and users with less than 5 ratings were removed. Also, both datasets were split into training, validation, and test sets with a  8:1:1 ratio. A detailed description of each dataset is given below:

\textbf{(1) MovieLens.} This includes user ratings on movies and is one of the most popular datasets in the recommendation field. The dataset was originally constructed by Grouplens\footnote{https://grouplens.org/datasets/movielens}. Other than user-item interactions, we leveraged features such as age and gender for users and genres for movies. In addition to the genre information  included in the MovieLens dataset, we collected and used the director and actor information of each movie from IMDB\footnote{https://www.imdb.com/interfaces/}. 

\textbf{(2) Weibo-Douban.} The dataset contains data from the Web services Weibo and Douban. The dataset was first introduced in \cite{yang2018social}. Weibo is a Twitter-like service used in China and Douban is a chinese review website for movies and books. The dataset contains user ratings on movies collected from Douban and user information from Weibo. The authors collected user tags and used them as user features. Douban also provides movie tags that are labeled by users. As there are approximately 50000 tags for items and users, we used only the 50 most frequently used tags for items and users respectively.

\begin{table}
  \label{tab:freq}
  \caption{Statistics of Datasets}
  \begin{tabular}{ccccc}
    \toprule
    Dataset&User&Item&Ratings&Sparsity\\
    \midrule
    MovieLens & 944 & 1,683 & 10,0000 & 93.70\%\\
    Weibo-Douban & 5,000 & 37,993 & 870,740 & 99.54\%\\

  \bottomrule
\end{tabular}
\vspace{-3.5mm}
\end{table}

\subsection{Baselines}
We compared our model with various LF models. We used the official implementation of the baseline models when possible. For the baseline models NFM and xDeepFM, we used Deep CTR's implementation\footnote{https://deepctr-doc.readthedocs.io/en/latest/index.html}.

\textbf{Biased MF \cite{koren2009matrix}.} This is the most basic MF model which uses only user-item interaction data, and user and item biases. Biased MF is widely used as a baseline in recommendation tasks, especially rating prediction tasks.

\textbf{NCF \cite{he2017neural}.} Neural collaborative filtering is a deep learning based MF model. Most of the deep learning approaches focus on improving the shallow structures of collaborative filtering models, which are similar to the structures of NCF models.

\textbf{AFM \cite{Chen2018heterogeneousattention}.} Attention-driven factor model is a state-of-the-art deep learning model that uses both content features and user-item feedback. AFM uses attention mechanism to model user different preferences on for item content features.

\textbf{LibFM \cite{rendle2010factorization}.} We used the official implementation of Factorization Machine (FM) released by the authors.

\textbf{NFM \cite{he2017nfm}.} This is one of the first deep neural network based FM models. The model is comprised of FM's linear structure and a non linear deep neural network structure; the model captures higher order feature interactions.  

\textbf{xDeepFM \cite{lian2018xdeepfm}.} State-of-the-art deep learning model that captures high-order feature interactions. We didn\textquotesingle t include other recent FM based deep learning models such as Wide\&Deep or Deep\&Cross as xDeepFM outperforms those models in various settings. 

\subsection{Implementation Details}
We implemented our model in PyTorch. We used an embedding dimension of 64 for all the MF models and our model. For the FM models, we used an embedding dimension of 16 as using a smaller embedding size is more effective. The Adam optimizer with a learning rate of 1e-3 and l2-regularization rate of 1e-4 is used for all the models. For the multi-head self-attention layer in SAIN, we used one self-attention layer and two attention heads. To prevent overfitting, we add a batch normalization layer after the self-attention layer, and apply dropout with a ratio of 0.1. Our source code is available at \url{https://github.com/seongjunyun/SAIN}.

\subsection{Performance Comparison}\label{subsec:perf_comp}
We summarized our experimental results in Table \ref{tab:result}. 
Our experimental results show that the MF models obtain higher performance on datasets with more user-item interactions (e.g. MovieLens dataset). The FM models obtained higher performance on the Weibo-Douban dataset, which is relatively sparser. In terms of RMSE, our model outperforms the best performing baseline model by 2.13\% and 1.01\% on MovieLens and Weibo-Douban, respectively. While considering complex feature interactions, our model SAIN can effectively learn feature 
representations and integrate content and feedback information. As a result, SAIN achieves the highest 
performance on both datasets.

\begin{table}[]
\caption{Experimental results on the MovieLens and Weibo-Douban datasets.}
\begin{tabular}{@{}|c||cc||cc|@{}}
\toprule
Dataset       & \multicolumn{2}{c||}{Movie-Lens}    & \multicolumn{2}{c|}{Weibo-Douban}  \\ \hline\hline
Models        & RMSE            & MAE             & RMSE            & MAE             \\ \hline\hline
BiasedMF      & 0.9132          & 0.7165          & 0.7933          & 0.6169               \\
NCF           & 0.9040          & 0.7150          & 0.7805          & 0.6158               \\
AFM           & 0.9126          & 0.7175          & 0.7717          & 0.6034               \\
LibFM            & 0.9234          & 0.7252          & 0.7755          & 0.6141               \\
NFM           & 0.9134          & 0.7189          & 0.7744          & 0.6111          \\
xDeepFM       & 0.9104          & 0.7155          & 0.7730          & 0.6077          \\ \hline\hline
\textbf{SAIN} & \textbf{0.8847} & \textbf{0.6958} & \textbf{0.7639} & \textbf{0.5957} \\ 
\bottomrule
\end{tabular}

\vspace{-5.8mm}
\label{tab:result}
\end{table}

\subsection{Analysis}

\subsubsection{Effect of Multi-head Self-attention}
\begin{figure}[t]
    \begin{center}
    \includegraphics[width=8.5cm,height=4cm]{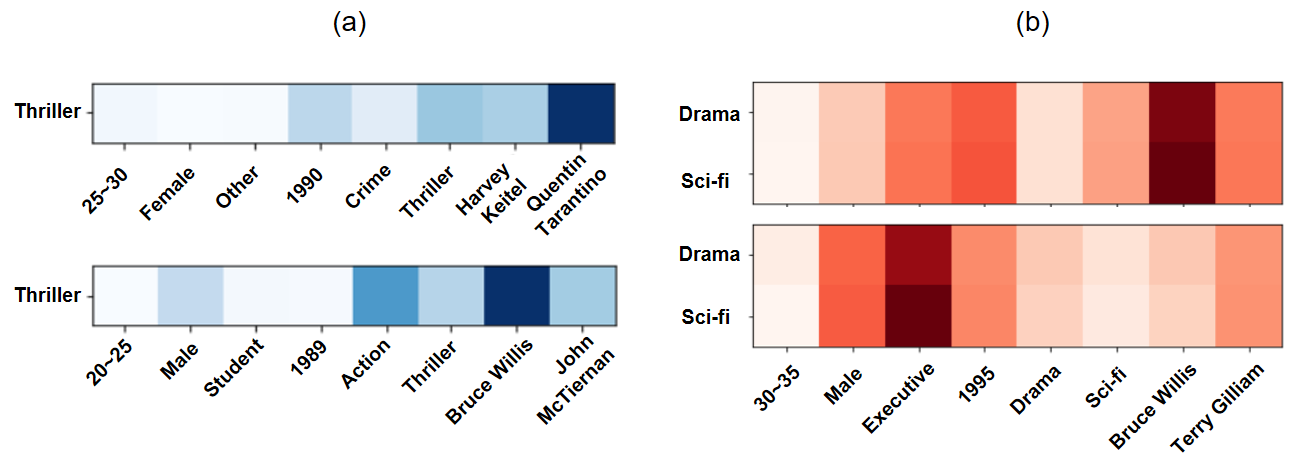}
    \caption{Attention score visualization. (a) Comparison of attention scores of two movies(same genre). (b) Shows the attention scores of two different heads for the same movie}
    \label{fig:attention vis}
    \end{center}
    \vspace{-5mm}
\end{figure}

As stated in \ref{subsec:feat_lev_layer}, we use a multi-head self attention structure in the feature-level interaction layer. We checked the attention scores of features to see how multiple features influence each other. The visualization of the attention scores is provided in Figure \ref{fig:attention vis}. Example (a) in Figure \ref{fig:attention vis} shows that two movies in the same thriller genre can have very different attention patterns. One movie focuses on an actor (Bruce Willis) and the other movie focuses on a director (Quentin Tarantino). Furthermore, we could verify that each head captures different aspects of feature interactions. Example (b) in Figure \ref{fig:attention vis} shows the attention score of a same movie can have very different patterns in different attention head. One focused on user features (gender, occupation), whereas the other one concentrated on item features (actor, director).

\subsubsection{Effect of Top-K features.} 
\begin{figure}[h]
    \begin{center}
    \includegraphics[width=5.8cm,height=2.3cm]{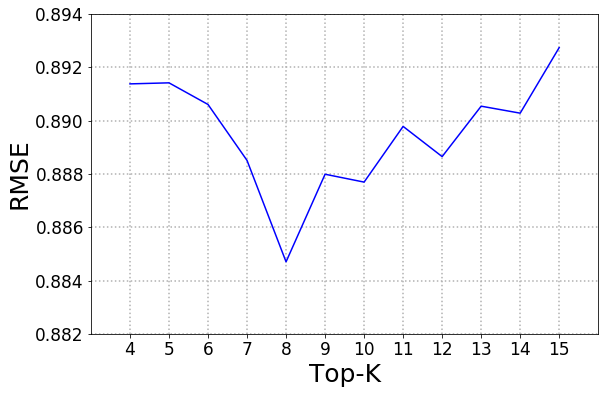}
    \caption{Changes in performance while varying K}
    \label{fig:top-k}
    \end{center}
    \vspace{-5.5mm}
\end{figure}

We used only features with the K highest attention scores to avoid overfitting. The changes in performance were observed while varying K. As demonstrated in Figure \ref{fig:top-k}, the prediction accuracy increases when K is 8. 

%% file: sections/Conclusion.tex
\section{Conclusion}
In this paper, we proposed Self-Attentive Integration Network (SAIN) which is a model that effectively combines user-item feedback information and auxiliary information for recommendation tasks. We conducted experiments on the well-known datasets MovieLens and Weibo-Douban. The experimental results show that our model outperforms the state-of-the-art LFM models. 